\documentclass[a4paper,12pt]{article}

\usepackage[T1]{fontenc}
\usepackage[utf8]{inputenc}
\usepackage{microtype}
\usepackage{amsmath,amssymb,amsthm,mathtools}

\usepackage{authblk}
\usepackage{graphicx}
\usepackage{tikz}
\usetikzlibrary{arrows.meta}
\usepackage{float}

\usepackage{enumitem}
\usepackage[font=small,labelfont=bf,labelsep=colon]{caption}
\usepackage{geometry}
\usepackage{xurl}
\usepackage{hyperref}
\hypersetup{hidelinks}

\usepackage[svgnames, table]{xcolor}
\usepackage[normalem]{ulem}

\usepackage{orcidlink}

\usepackage{flafter}
\usepackage{placeins}
\usepackage{cite} 

\title{Graph construction in QUBO-based recursive phylogenetic tree reconstruction}

\author[1,2]{Kanazawa Yoshiki \orcidlink{0009-0007-6007-553X}}
\author[2,3]{Takahiko Koyama \orcidlink{0000-0003-1694-9061}}

\affil[1]{Graduate School of Media and Governance, Keio University, Fujisawa, Kanagawa, Japan}
\affil[2]{Human Biology-Microbiome-Quantum Research Center (WPI-Bio2Q), Keio University, Tokyo, Japan}
\affil[3]{Keio University Sustainable Quantum Artificial Intelligence Center (KSQAIC), Keio University, Tokyo, Japan}
\date{\vspace{-1em}\small
\texttt{ky01752221@keio.jp}, \texttt{takahiko.koyama@keio.jp}
}

\begin{document}
\maketitle

\begin{abstract}
Molecular sequence data are used to reconstruct evolutionary relationships among taxa, but reconstruction accuracy depends not only on the tree-building method but also on how pairwise sequence relationships are represented.
In graph-based approaches, this representation determines the edge weights used during graph partitioning.
Here, we evaluated sequence-to-affinity representations in a recursive normalized-cut (Ncut) framework whose graph-partitioning subproblems were formulated as quadratic unconstrained binary optimization (QUBO) models and solved using Simulated Bifurcation.
The analysis used simulated amino-acid and nucleotide datasets spanning multiple tree-generation settings and levels of evolutionary divergence.
We compared normalized bit-score affinities with representations derived from transformed sequence similarities and evolutionary distances, examined cardinality-preserving post-swap refinement, and used neighbor joining (NJ) as a distance-based comparator.

Affinity representation substantially affected internal split recovery, particularly for nucleotide data.
JC69-based local affinities maintained comparatively high accuracy as divergence increased, whereas normalized bit-score and BLAST-derived kernel representations declined more markedly.
Post-swap refinement generally improved recovery, but improvements were not consistent across individual reconstructions.
NJ achieved higher mean split recovery than corresponding affinity-based recursive Ncut reconstructions for WAG and JC69 distances across all evaluated conditions, whereas recursive Ncut outperformed NJ for BLAST-derived logarithmic distances under some conditions.

These results show that graph construction is an important determinant of recursive Ncut-based phylogenetic reconstruction, but a representation that performs well within Ncut does not necessarily provide the most accurate use of the underlying pairwise distances.
Pairwise representation, affinity transformation, optimization, and recursive tree construction should therefore be evaluated jointly.
For biological applications, these findings emphasize that the computational representation of sequence similarity and evolutionary divergence is itself an important component of phylogenetic inference.
\end{abstract}

\begin{figure*}[t]
    \centering

    \includegraphics[width=\textwidth]
    {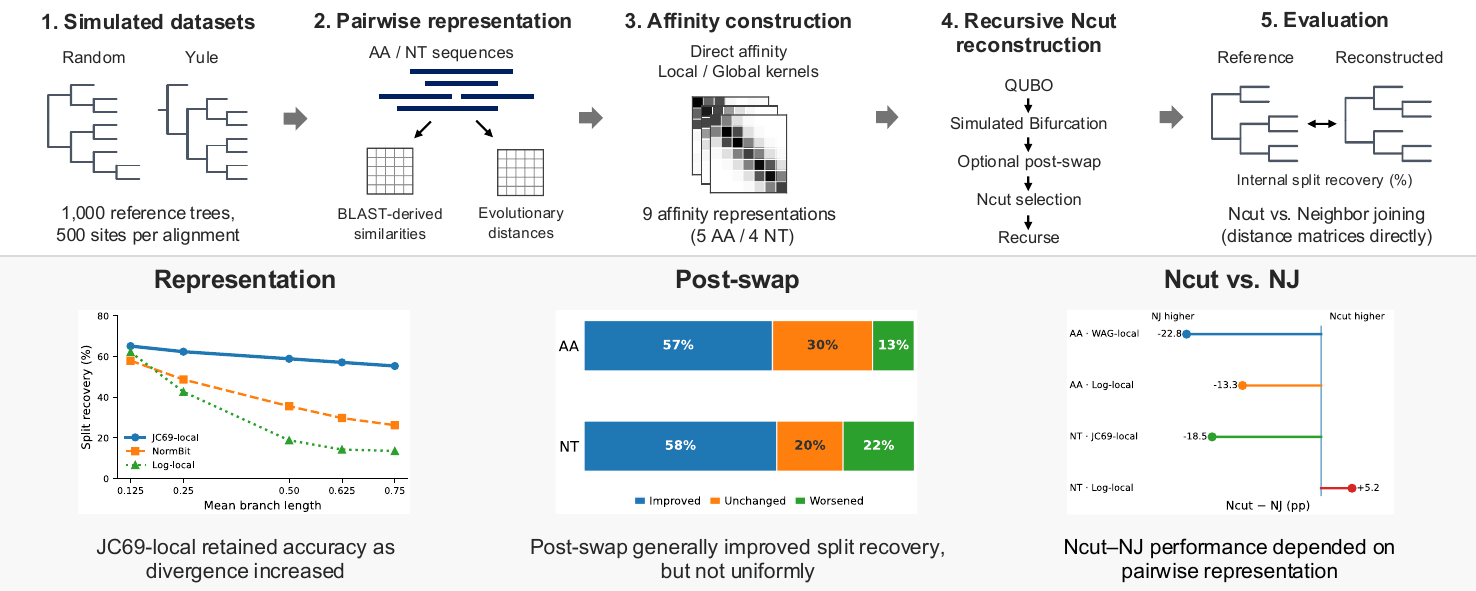}

    \caption*{\textbf{Graphical abstract.} Overview of the study workflow and main findings.}
\end{figure*}
\FloatBarrier

\section{Introduction}
Molecular sequence data provide a record of evolutionary divergence, but translating sequence variation into representations that enable accurate reconstruction of evolutionary relationships remains a central challenge.
Phylogenetic trees can be reconstructed using maximum-likelihood, Bayesian, and pairwise-distance-based approaches~\cite{Felsenstein1981,Rannala1996,Saitou1987}.
In pairwise-distance-based methods such as neighbor joining (NJ), trees are constructed directly from a matrix of pairwise evolutionary distances~\cite{Saitou1987}.
Reconstruction therefore depends not only on the tree-building algorithm but also on how pairwise sequence relationships are represented.
This choice becomes particularly important when the resulting matrix directly defines the edge weights of a graph used in an optimization or clustering procedure.

Pairwise sequence relationships can also be represented as a weighted graph and recursively partitioned to construct a tree~\cite{Matsui2020}.
A minimum-cut approach partitions a graph by minimizing the total weight of edges between the resulting groups, but this criterion can favor highly unbalanced partitions~\cite{Wu1993,Shi2000}.
The normalized-cut (Ncut) criterion instead accounts for each partition's association with the full graph~\cite{Shi2000}.
Graph-partitioning subproblems can also be formulated as quadratic unconstrained binary optimization (QUBO) models, an approach recently explored in phylogenetic applications~\cite{Dinneen2023,Zhang2026}.
The Normalized-Minimum cut by Digital Annealer (NMcutDA) method applied this strategy to recursive Ncut-based phylogenetic reconstruction using amino-acid similarity graphs and the Fujitsu Digital Annealer~\cite{Onodera2023}; its recursive Ncut framework and normalized bit-score representation serve as the baseline for the present study.

Despite this demonstrated feasibility, the performance of NMcutDA varied substantially across simulation conditions.
Reconstruction accuracy was relatively low for datasets generated from trees with short mean branch lengths, and lower Ncut values did not necessarily correspond to more accurate reconstructed trees~\cite{Onodera2023}.
These results indicate that further optimization of the graph-partitioning objective alone may not be sufficient to improve phylogenetic accuracy.
Because Ncut is evaluated on a weighted graph derived from pairwise sequence relationships, the selected partitions may also depend on how those relationships are converted into edge weights.
Different pairwise representations and distance-to-affinity transformations can induce different relative edge weights for the same sequence pairs.
However, the effect of such graph-construction choices has not been systematically evaluated within a recursive Ncut-based phylogenetic reconstruction framework.

In this study, we investigated how graph construction influences recursive Ncut-based phylogenetic reconstruction.
Using NMcutDA as a methodological baseline, we compared its BLAST-derived normalized bit-score representation with affinity graphs derived from transformed sequence similarities and model-corrected evolutionary distances~\cite{Altschul1990,Dufour2010,Onodera2023}.
QUBO subproblems were solved using Simulated Bifurcation (SB)~\cite{Goto2019,Goto2021}.
We further compared recursive Ncut with NJ applied directly to the corresponding distance matrices and examined whether Ncut-reducing post-swap refinement improved phylogenetic accuracy.
These analyses were performed on simulated amino-acid and nucleotide datasets spanning multiple tree-generation settings and levels of evolutionary divergence.
More broadly, this study examines how computational representations of sequence similarity and evolutionary divergence influence phylogenetic reconstruction from molecular sequence data.

\section{Materials and methods}
\label{sec:materials_methods}
\subsection{Overview}

We evaluated a recursive Ncut-based phylogenetic reconstruction framework using simulated amino-acid and nucleotide sequence datasets. The analysis consisted of four main steps:

\begin{enumerate}[
    label=(\roman*),
    leftmargin=*,
    itemsep=2pt,
    topsep=4pt,
    parsep=0pt
]
    \item generation of reference trees and simulation of sequence alignments,
    \item construction of pairwise sequence-similarity or evolutionary-distance representations,
    \item conversion of these representations into affinity matrices, and
    \item recursive tree reconstruction by Ncut-based graph partitioning.
\end{enumerate}

The binary graph-partitioning subproblems were formulated as QUBO models and solved using SB.

As an external comparator, NJ was applied directly to the corresponding pairwise distance matrices, without conversion to affinity matrices. Trees reconstructed using the recursive Ncut framework and NJ were compared with the corresponding reference trees using internal split recovery.

\subsection{Simulation of reference trees and sequence alignments}
Reference trees were generated under two tree-generation settings. Random bifurcating tree topologies were generated using the \texttt{rtree} function in the \texttt{ape} package (version 5.8-1)~\cite{Paradis2019}, whereas Yule trees were generated under a pure-birth process using the \texttt{TreeSim} package (version 2.4)~\cite{Stadler2011}. All reference trees contained $n=30$ taxa.

To examine the effect of evolutionary divergence, we considered five target mean branch lengths, measured in substitutions per site:
\[
0.125,\ 0.250,\ 0.500,\ 0.625,\ 0.750.
\]
For random-tree datasets, individual branch lengths were independently sampled from an exponential distribution with the specified mean.
For Yule-tree datasets, the branch lengths generated under the pure-birth process were rescaled so that their mean matched the specified target value.
For each combination of tree-generation setting and target mean branch length, 100 independent reference trees were generated.
This yielded 1,000 reference trees in total. On each reference tree, one amino-acid alignment and one nucleotide alignment, each comprising 500 sites, were simulated.

Amino-acid sequences were simulated under the WAG substitution model~\cite{Whelan2001}, with among-site rate heterogeneity represented by a discrete gamma distribution with five rate categories and shape parameter $\alpha=1$.
Nucleotide sequences were simulated under the JC69 substitution model~\cite{Jukes1969}, using the same discrete gamma-rate setting.
Sequence simulation was performed using AliSim in IQ-TREE version 3.1.1~\cite{Nguyen2015,LyTrong2022}.

\subsection{BLAST-based similarity matrices}
For each simulated alignment, all-against-all sequence comparisons were performed using BLAST+ version 2.17.0+~\cite{Camacho2009}.
Amino-acid datasets were processed using BLASTP, whereas nucleotide datasets were processed using BLASTN.
The E-value threshold was set to 10, and at most one high-scoring segment pair was retained for each query--subject pair.
Pairs for which no BLAST hit was reported were assigned a bit score of zero.

Let $b_{ij}^{\mathrm{dir}}$ denote the maximum bit score reported when sequence $i$ was used as the query and sequence $j$ as the subject.
The directional bit-score matrix was symmetrized by arithmetic averaging,
\[
b_{ij}
=
\frac{1}{2}
\left(
b_{ij}^{\mathrm{dir}} + b_{ji}^{\mathrm{dir}}
\right).
\]
Following the normalized bit-score representation used in previous sequence-similarity analyses~\cite{Dufour2010,Onodera2023}, we defined
\[
s_{ij}
=
\frac{b_{ij}}
{\frac{1}{2}(b_{ii}+b_{jj})}.
\]
The normalized similarities were restricted to the interval $[0,1]$ and used as the baseline sequence-similarity representation.
For graph construction, diagonal affinity values were set to zero.

We also converted the normalized bit-score similarities into a logarithmic distance matrix,
\[
d_{ij}
=
-\log\left(\max(s_{ij},\varepsilon)\right),
\]
where $\varepsilon=10^{-12}$ was used for numerical stability and $d_{ii}=0$.

Two kernel transformations were applied to the logarithmic distance matrix.
For the global Gaussian kernel, the affinity was defined as
\[
K_{ij}
=
\exp\left(
-\frac{d_{ij}^{2}}{\lambda\sigma^{2}}
\right),
\]
where $\sigma$ was set to the median of the positive off-diagonal distances and $\lambda=1$.
For the self-tuning local kernel~\cite{ZelnikManor2004}, the local scale parameter $\sigma_i$ was defined as the $k$-th nearest positive distance from sequence $i$.
We used $k=7$.
The affinity was then defined as
\[
K_{ij}
=
\exp\left(
-\frac{d_{ij}^{2}}{\lambda\sigma_i\sigma_j}
\right),
\]
with $\lambda=1$.
Diagonal affinity values were set to zero for both kernel-derived matrices.

The logarithmic distance matrix was used either to construct affinity matrices for recursive Ncut-based reconstruction or directly as input to neighbor joining, as described below.

\subsection{Evolutionary-distance matrices}
In addition to BLAST-derived similarities, evolutionary-distance matrices were constructed directly from the simulated alignments.

For amino-acid datasets, the pairwise amino-acid $p$-distance was computed as the fraction of mismatched sites among positions containing standard amino-acid states in both sequences.
Sites containing gaps or non-standard residues in either sequence were excluded from the pairwise comparison.
As a simple symmetric reference model, we defined a 20-state equal-rates correction in which all amino-acid states were treated equivalently.
The corrected distance was defined as
\[
d_{ij}^{\mathrm{ER20}}
=
-\frac{19}{20}
\log\left(
1-\frac{20}{19}p_{ij}
\right).
\]
For numerical stability, $p_{ij}$ was restricted to the interval
\[
0 \leq p_{ij} \leq \frac{19}{20}-\varepsilon,
\]
where $\varepsilon=10^{-12}$.

We also computed a pairwise maximum-likelihood distance under the WAG amino-acid substitution model~\cite{Whelan2001}.
The WAG exchangeability parameters and equilibrium amino-acid frequencies were used to construct the instantaneous rate matrix, which was normalized to an expected substitution rate of one.
For each pair of amino-acid sequences, transition probabilities were evaluated over 501 equally spaced evolutionary times from $t=0$ to $t=5$.
The pairwise log-likelihood was calculated from the observed amino-acid pair counts, and the value of $t$ that maximized the log-likelihood was used as the WAG-based distance.
The WAG distance calculation did not incorporate among-site gamma-rate averaging.

For nucleotide datasets, the pairwise nucleotide $p$-distance was computed as the fraction of mismatched sites among positions containing one of the four standard nucleotide states in both sequences.
Sites containing ambiguous bases or gaps were excluded from the pairwise comparison.
The JC69-corrected distance was defined as~\cite{Jukes1969}
\[
d_{ij}^{\mathrm{JC69}}
=
-\frac{3}{4}
\log\left(
1-\frac{4}{3}p_{ij}
\right).
\]
For numerical stability, $p_{ij}$ was restricted to
\[
0 \leq p_{ij} \leq \frac{3}{4}-\varepsilon,
\]
using the same value $\varepsilon=10^{-12}$.

For recursive Ncut-based reconstruction, all evolutionary-distance matrices were converted into affinity matrices using the self-tuning local kernel described above. The untransformed distance matrices were also used directly for neighbor-joining reconstruction.

\subsection{Evaluated affinity representations}
For amino-acid datasets, five sequence-to-affinity representations were evaluated:
\begin{description}[style=nextline, leftmargin=0pt, itemsep=2pt, topsep=2pt]
    \item[NormBit]
    the normalized bit-score similarity matrix used directly as the affinity matrix.

    \item[Log-local]
    the self-tuning local kernel applied to the BLAST-derived logarithmic distance.

    \item[Log-global]
    the global Gaussian kernel applied to the BLAST-derived logarithmic distance.

    \item[ER20-local]
    the self-tuning local kernel applied to the ER20-corrected amino-acid distance.

    \item[WAG-local]
    the self-tuning local kernel applied to the WAG-based pairwise maximum-likelihood distance.
\end{description}

For nucleotide datasets, four sequence-to-affinity representations were evaluated:
\begin{description}[style=nextline, leftmargin=0pt, itemsep=2pt, topsep=2pt]
    \item[NormBit]
    the normalized bit-score similarity matrix used directly as the affinity matrix.

    \item[Log-local]
    the self-tuning local kernel applied to the BLAST-derived logarithmic distance.

    \item[Log-global]
    the global Gaussian kernel applied to the BLAST-derived logarithmic distance.

    \item[JC69-local]
    the self-tuning local kernel applied to the JC69-corrected nucleotide distance.
\end{description}

Each representation was evaluated within the same recursive Ncut-based reconstruction framework, both with and without post-swap refinement.

\subsection{Normalized-cut-based recursive tree reconstruction}
Given an affinity matrix $W=(w_{ij})$, we reconstructed a binary tree by recursively dividing the taxa into two nonempty subsets.
At each step, graph partitioning was performed using only the affinity values among taxa in the current subset.
For a bipartition $(A,B)$ of the current taxon set $V$, the cut value was defined as
\[
\operatorname{cut}(A,B)
=
\sum_{i\in A,\,j\in B} w_{ij}.
\]
The association of a subset $A$ with $V$ was defined as
\[
\operatorname{assoc}(A,V)
=
\sum_{i\in A,\,j\in V} w_{ij}.
\]
The normalized cut value was then defined as~\cite{Shi2000}
\[
\operatorname{Ncut}(A,B)
=
\frac{\operatorname{cut}(A,B)}
{\operatorname{assoc}(A,V)}
+
\frac{\operatorname{cut}(A,B)}
{\operatorname{assoc}(B,V)}.
\]

Following the cardinality-wise graph-cut strategy used in NMcutDA~\cite{Onodera2023}, candidate bipartitions were generated by solving cardinality-constrained minimum-cut problems.
For a fixed target cardinality $c$, binary variables $x_i\in\{0,1\}$ were introduced, where $x_i=1$ indicates membership in one side of the partition.
The minimum-cut problem was represented by the quadratic unconstrained binary optimization objective
\[
\sum_{i<j} w_{ij}(x_i-x_j)^2
+
\alpha
\left(
\sum_i x_i-c
\right)^2.
\]
The first term represents the cut value, whereas the second term penalizes deviations from the target cardinality.

For a current node containing $m$ taxa, target cardinalities
\[
c=1,\ldots,\left\lfloor\frac{m}{2}\right\rfloor
\]
were considered.
The QUBO problem corresponding to each target cardinality was solved using the SB algorithm under the computational settings described below.
The initial cardinality-penalty coefficient was set to $\alpha=128$.
If no solution satisfying the target cardinality was obtained, the coefficient was doubled and the QUBO problem was solved again, for a maximum of six attempts per target cardinality.

Among the binary solutions returned by the SB algorithm, solutions satisfying the target cardinality were identified. For each target cardinality, the feasible solution with the smallest minimum-cut objective value was retained as the candidate bipartition.
The Ncut value of each retained candidate was then calculated after optimization. Among the candidates obtained across all target cardinalities, the bipartition with the smallest Ncut value was selected. The two resulting subsets were recursively partitioned until singleton leaves were obtained.

If no valid nontrivial bipartition was obtained from the binary optimization procedure, a deterministic spectral fallback was applied~\cite{vonLuxburg2007}. Let $D$ denote the diagonal degree matrix of the current affinity graph.
The symmetric normalized graph Laplacian was defined as
\[
L_{\mathrm{sym}}
=
D^{-1/2}(D-W)D^{-1/2}.
\]
Taxa were ordered according to the eigenvector associated with the second-smallest eigenvalue of $L_{\mathrm{sym}}$ and divided into two subsets at $\lfloor m/2\rfloor$.

\subsection{Post-swap refinement}
For each affinity setting, reconstruction was evaluated both without and with cardinality-preserving Ncut-based post-swap refinement.
When post-swap refinement was enabled, each feasible bipartition obtained for a target cardinality was refined before its Ncut value was used for candidate selection.
At each refinement pass, all possible swaps involving one taxon from each subset were evaluated.
Among swaps that reduced the Ncut value by more than a tolerance of $10^{-12}$, the swap resulting in the lowest Ncut value was applied.
Because each swap exchanged one taxon between the two subsets, the sizes of the two subsets remained unchanged.
The procedure was repeated until no improving swap was found or a maximum of 20 passes was reached.
Thus, each accepted swap strictly reduced the Ncut value of the candidate bipartition being refined.

\subsection{Neighbor-joining reconstruction}
To provide a standard distance-based comparator, NJ trees were reconstructed directly from the pairwise distance matrices before their conversion into affinity matrices.

For amino-acid datasets, NJ was applied to the BLAST-derived logarithmic distance, the ER20-corrected distance, and the WAG-based distance. For nucleotide datasets, NJ was applied to the BLAST-derived logarithmic distance and the JC69-corrected distance.
Because the Log-local and Log-global representations differed only in the subsequent distance-to-affinity transformation, the same NJ tree constructed from the BLAST-derived logarithmic distance was used for comparison with both representations.

Trees were reconstructed using the \texttt{nj} function in the \texttt{ape} package version 5.8-1 under R version 4.3.2. No affinity-kernel transformation or post-swap refinement was applied to the NJ reconstructions.

\subsection{Evaluation of reconstructed trees}
Each tree reconstructed using either the recursive Ncut framework or neighbor joining was compared with its corresponding reference tree using nontrivial internal bipartitions.
A split was considered nontrivial when both sides of the bipartition contained at least two taxa.
Let $\mathcal{S}_{\mathrm{ref}}$ denote the set of nontrivial internal splits in the reference tree and $\mathcal{S}_{\mathrm{rec}}$ the corresponding set in the reconstructed tree.
Split recovery was defined as
\[
\operatorname{Recovery}(\%)
=
100
\frac{
|\mathcal{S}_{\mathrm{ref}}\cap\mathcal{S}_{\mathrm{rec}}|
}{
|\mathcal{S}_{\mathrm{ref}}|
}.
\]
This measure represents the percentage of nontrivial internal splits in the reference tree that were recovered in the reconstructed tree.
For fully bifurcating trees defined on the same taxon set, split recovery is related to the normalized Robinson--Foulds split distance by
\[
\operatorname{Recovery}(\%)
=
100\left(1-d_{\mathrm{RF}}^{\mathrm{norm}}\right)
\]~\cite{Robinson1981}.

For the recursive Ncut framework, results were summarized over the 100 replicate datasets for each combination of sequence type, tree-generation setting, target mean branch length, affinity representation, and post-swap mode.
For neighbor joining, results were summarized over the 100 replicate datasets for each combination of sequence type, tree-generation setting, target mean branch length, and pairwise distance representation.

\subsection{Statistical analysis}
For each affinity representation, mean split recovery and its 95\% confidence interval were estimated separately within each combination of sequence type, tree-generation setting, target mean branch length, and post-swap mode.

For direct comparisons with NormBit, each transformed affinity representation was paired with NormBit for the same reference tree and simulated sequence alignment.
The paired difference was calculated as the split recovery of the transformed representation minus that of NormBit, such that positive values indicated higher recovery for the transformed representation.
Paired differences were summarized within each condition and across conditions.

The effect of post-swap refinement was evaluated using paired results obtained from the same reference tree and simulated sequence alignment.
Reconstructions performed with and without post-swap refinement were paired for each dataset and affinity representation.
The paired difference was defined as
\[
\Delta\operatorname{Recovery}_{\mathrm{postswap}}
=
\operatorname{Recovery}_{\mathrm{with\ postswap}}
-
\operatorname{Recovery}_{\mathrm{without\ postswap}}.
\]

To assess the consistency of the post-swap effect across individual reconstructions, paired outcomes were classified as improved, unchanged, or worsened according to whether $\Delta\operatorname{Recovery}_{\mathrm{postswap}}$ was positive, zero, or negative, respectively. Counts and proportions in these three categories were summarized for each affinity representation and across representations, separately for amino-acid and nucleotide datasets. All paired reconstructions were included in this summary, including cases in which no post-swap exchange was accepted.

For comparisons with neighbor joining, recursive Ncut and NJ results were paired by sequence type, tree-generation setting, target mean branch length, replicate, and underlying pairwise distance representation.
ER20-local, WAG-local, and JC69-local were compared with NJ trees constructed from the corresponding ER20, WAG, and JC69 distance matrices, respectively.
Both Log-local and Log-global were compared with the NJ tree constructed from the same BLAST-derived logarithmic distance matrix.
The primary comparisons with NJ used recursive Ncut results obtained without post-swap refinement.

For the comparison between recursive Ncut and NJ, the paired difference in split recovery was defined as
\[
\Delta\operatorname{Recovery}_{\mathrm{Ncut-NJ}}
=
\operatorname{Recovery}_{\mathrm{Ncut}}
-
\operatorname{Recovery}_{\mathrm{NJ}}.
\]
Differences were reported in percentage points, with positive values indicating higher split recovery by recursive Ncut reconstruction.

Mean split-recovery values and mean paired differences were accompanied by 95\% confidence intervals estimated using nonparametric bootstrap resampling.
For summaries of individual affinity representations, the 100 replicate datasets were resampled within each combination of sequence type, tree-generation setting, target mean branch length, affinity representation, and post-swap mode.
For each condition-specific paired comparison, the 100 paired observations were resampled as paired units within the corresponding combination of sequence type, tree-generation setting, target mean branch length, and the relevant affinity representation or distance comparison.
For overall Ncut--NJ summaries combining tree-generation and branch-length conditions, bootstrap resampling was performed separately within each tree-generation-setting-by-branch-length stratum, and the resulting stratum means were averaged with equal weight.
Confidence intervals were calculated from 10,000 bootstrap samples using the percentile method.

\subsection{Computational implementation}
The analysis workflow was implemented primarily in Python, with reference-tree generation performed in R version 4.3.2.
QUBO optimization was performed using discrete Simulated Bifurcation~\cite{Goto2021}, as implemented in the \texttt{simulated-bifurcation} package version 2.0.0~\cite{Ageron2025}. 
The package was executed using PyTorch version 2.13.0~\cite{Paszke2019} with CUDA support.

Simulated Bifurcation optimization was performed in discrete mode using 256 agents, a maximum of 5,000 optimization steps, and double-precision floating-point arithmetic. Heating and early stopping were disabled.
Before optimization, each QUBO matrix was divided by its maximum absolute coefficient.
Candidate objective values were subsequently recalculated using the original, unscaled QUBO coefficients.
The same implementation and optimization settings were used across all affinity representations, sequence types, and replicate datasets.

A global random seed of 20260416 was used to derive deterministic solver seeds for individual QUBO optimization runs. Each solver seed was determined from the dataset identifier, recursive node, node size, target cardinality, and optimization-attempt number.
The same global seed was also used to generate deterministic per-dataset seeds for reference-tree generation and sequence simulation.
The tree-generation seed was supplied to the corresponding R procedure, whereas the sequence-simulation seed was supplied to AliSim.

\section{Results}
\subsection{Amino-acid reconstruction accuracy varied across affinity representations}
\begin{figure*}[t]
    \centering

    \includegraphics[width=0.485\textwidth]
    {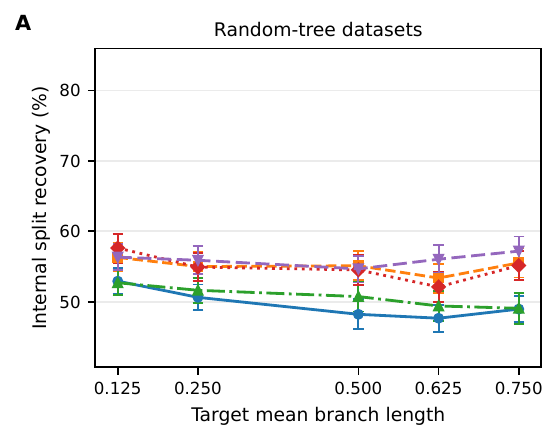}
    \hfill
    \includegraphics[width=0.485\textwidth]
    {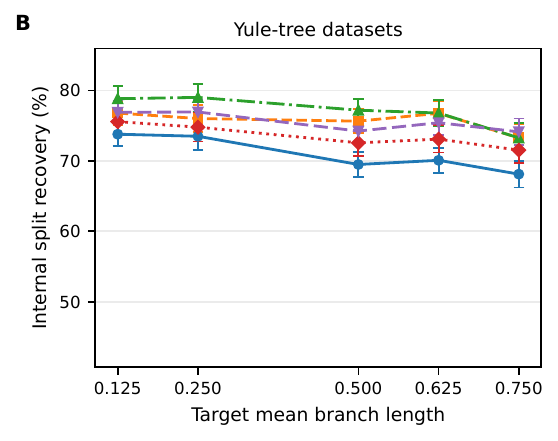}

    \par\vspace{-0.4em}

    \includegraphics[width=0.72\textwidth]
    {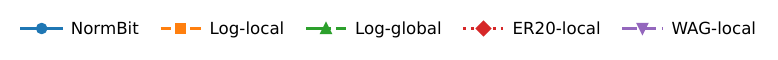}

    \caption{
    Amino-acid tree-reconstruction accuracy across affinity representations without post-swap refinement.
    \textbf{(A)} Random-tree datasets.
    \textbf{(B)} Yule-tree datasets.
    Points indicate mean internal split recovery across 100 replicate datasets for each target mean branch length, and error bars indicate nonparametric bootstrap 95\% confidence intervals.
    }
    \label{fig:aa_affinity}
\end{figure*}

To examine the effect of sequence-to-affinity representation independently of post-swap refinement, we first compared amino-acid reconstructions obtained without post-swap refinement (Fig.~\ref{fig:aa_affinity}).

For random-tree datasets, Log-local, ER20-local, and WAG-local yielded higher mean split recovery than NormBit at all five target mean branch lengths.
Their paired mean differences from NormBit ranged from 3.4 to 8.4 percentage points.
Log-global remained close to NormBit, yielding slightly lower mean recovery at a target mean branch length of 0.125 and nearly identical recovery at 0.750.
Across all five affinity representations and branch-length settings, mean split recovery ranged from 47.7\% to 57.6\%.
No clear monotonic relationship was observed between target mean branch length and reconstruction accuracy.

Split recovery was higher overall for Yule-tree datasets, ranging from 68.1\% to 79.0\%.
Log-global achieved the highest mean recovery at target mean branch lengths from 0.125 to 0.625, whereas WAG-local achieved the highest mean recovery at 0.750.
All four transformed representations yielded higher mean recovery than NormBit at each of the five branch-length settings, although the magnitude of improvement varied among representations and conditions.
The paired mean differences from NormBit ranged from 1.3 to 7.7 percentage points.

Across the 1,000 paired amino-acid reconstructions available for each transformed representation, the two largest overall mean paired differences relative to NormBit were 5.0 percentage points for Log-local and 5.4 percentage points for WAG-local.
Both representations yielded higher condition-level mean recovery than NormBit for every tree-generation setting and target mean branch length.

The affinity representation with the highest mean recovery depended on the tree-generation and evolutionary settings.
WAG-local showed the highest overall mean recovery across the amino-acid datasets, whereas Log-global performed best in most Yule-tree conditions.
Overall, transforming pairwise sequence information into kernel-based affinities generally improved mean reconstruction accuracy relative to the direct use of normalized bit scores, but the magnitude and consistency of the improvement were condition dependent.

\FloatBarrier

\subsection{JC69-based affinities achieved the highest nucleotide accuracy within the recursive Ncut framework}
\begin{figure*}[t]
    \centering

    \includegraphics[width=0.485\textwidth]
    {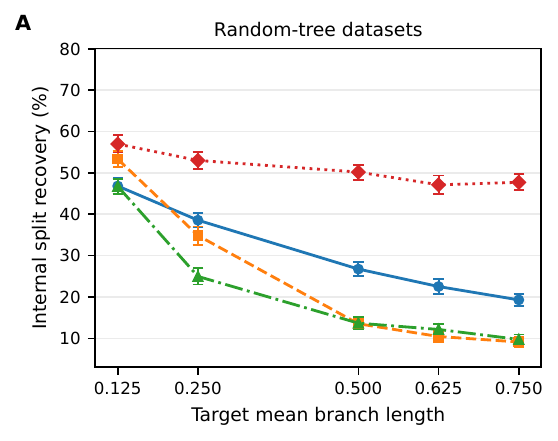}
    \hfill
    \includegraphics[width=0.485\textwidth]
    {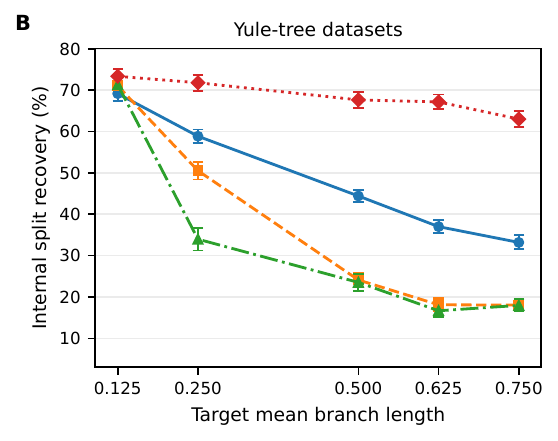}

    \par\vspace{-0.4em}

    \includegraphics[width=0.62\textwidth]
    {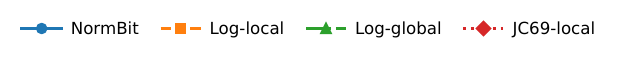}

    \caption{
    Nucleotide tree-reconstruction accuracy across affinity representations without post-swap refinement.
    \textbf{(A)} Random-tree datasets.
    \textbf{(B)} Yule-tree datasets.
    Points indicate mean internal split recovery across 100 replicate datasets for each target mean branch length, and error bars indicate nonparametric bootstrap 95\% confidence intervals.
    }
    \label{fig:nt_affinity}
\end{figure*}

We next evaluated the effect of affinity representation on nucleotide tree reconstruction within the recursive Ncut framework, without post-swap refinement (Fig.~\ref{fig:nt_affinity}).

For random-tree datasets, JC69-local yielded the highest mean split recovery at all five target mean branch lengths.
Its mean recovery ranged from 47.1\% to 57.0\%, whereas recovery for NormBit decreased from 46.8\% at a target mean branch length of 0.125 to 19.3\% at 0.750.
The paired mean improvement of JC69-local over NormBit ranged from 10.1 to 28.4 percentage points.

The two BLAST-derived kernel representations showed a different pattern. 
At a target mean branch length of 0.125, Log-local yielded a mean recovery of 53.3\%, compared with 46.8\% for NormBit, whereas Log-global produced nearly identical recovery at 46.7\%.
At target mean branch lengths from 0.250 to 0.750, however, both BLAST-derived kernel representations yielded lower mean recovery than NormBit.
Their recovery declined to below 10\% at a target mean branch length of 0.750.

The same general pattern was observed for Yule-tree datasets.
JC69-local achieved the highest mean recovery at all five target mean branch lengths, ranging from 63.0\% to 73.3\%.
Its paired mean improvement over NormBit ranged from 4.2 to 30.1 percentage points.
Log-local and Log-global both yielded higher mean recovery than NormBit at a target mean branch length of 0.125, but lower recovery at each of the four longer branch-length settings.

Across all 1,000 paired nucleotide reconstructions, JC69-local yielded higher split recovery than NormBit in 904 cases, identical recovery in 41 cases, and lower recovery in 55 cases.
The overall mean paired difference was 20.1 percentage points in favor of JC69-local (95\% CI, 19.5--20.8 percentage points).

Overall, nucleotide reconstruction accuracy showed a strong dependence on the pairwise representation.
JC69-local retained comparatively high recovery as the target mean branch length increased, whereas the direct NormBit representation and the BLAST-derived kernel representations showed substantially larger declines.

\FloatBarrier

\subsection{Post-swap refinement generally improved split recovery}
\begin{figure*}[t]
    \centering

    \includegraphics[width=0.485\textwidth]
    {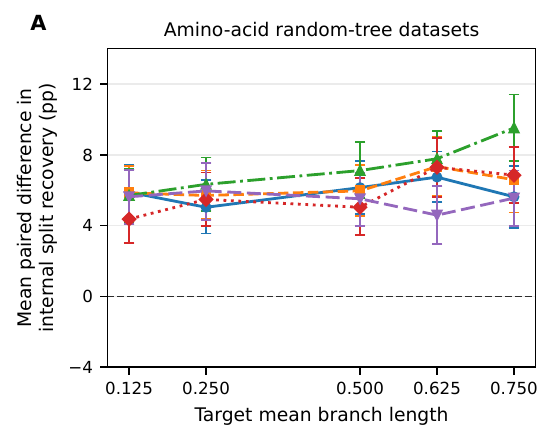}
    \hfill
    \includegraphics[width=0.485\textwidth]
    {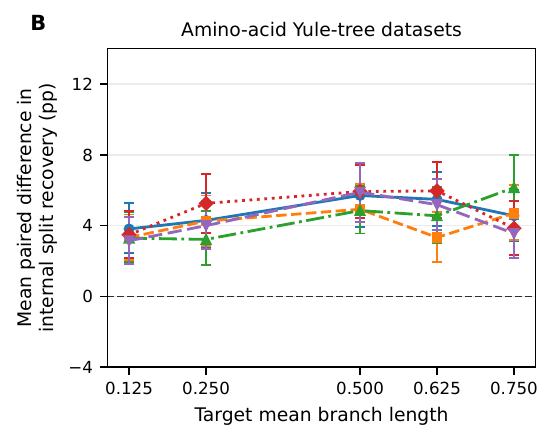}

    \par\vspace{0.3em}

    \includegraphics[width=0.485\textwidth]
    {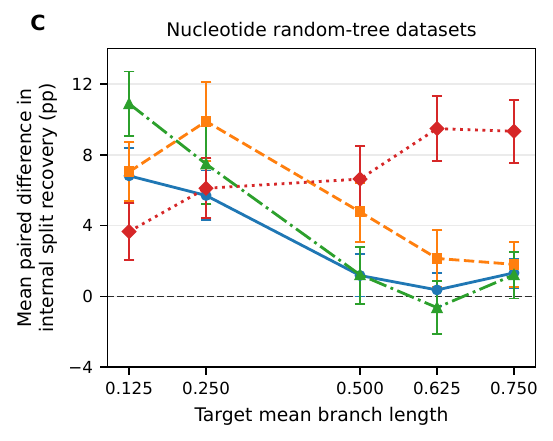}
    \hfill
    \includegraphics[width=0.485\textwidth]
    {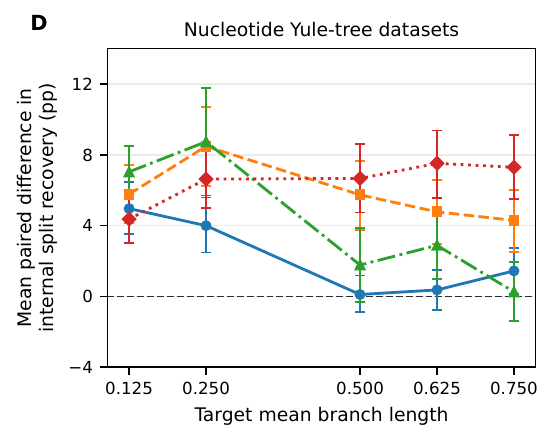}

    \par\vspace{-0.3em}

    \includegraphics[width=0.72\textwidth]
    {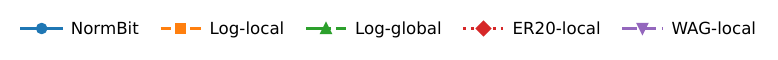}

    \par\vspace{-0.2em}

    \includegraphics[width=0.58\textwidth]
    {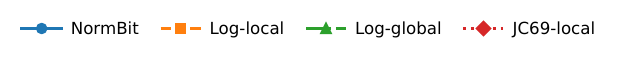}

    \caption{
    Effect of post-swap refinement on tree-reconstruction accuracy.
    \textbf{(A)} Amino-acid random-tree datasets.
    \textbf{(B)} Amino-acid Yule-tree datasets.
    \textbf{(C)} Nucleotide random-tree datasets.
    \textbf{(D)} Nucleotide Yule-tree datasets.
    Differences were calculated as split recovery with post-swap refinement minus split recovery without refinement for the same simulated dataset and affinity representation.
    Points indicate mean paired differences across 100 paired reconstructions per condition, and error bars indicate nonparametric bootstrap 95\% confidence intervals.
    Positive values indicate higher split recovery after post-swap refinement.
    }
    \label{fig:postswap}
\end{figure*}

We next evaluated the effect of cardinality-preserving post-swap refinement by comparing reconstructions obtained with and without refinement for the same simulated dataset and affinity representation (Fig.~\ref{fig:postswap}).

For amino-acid datasets, post-swap refinement increased mean split recovery in every combination of tree-generation setting, target mean branch length, and affinity representation.
For random-tree datasets, the mean paired improvement ranged from 4.4 to 9.5 percentage points.
The largest improvement was observed for Log-global at a target mean branch length of 0.750.
For Yule-tree datasets, the improvements were generally smaller, ranging from 3.1 to 6.1 percentage points.
The bootstrap confidence intervals were above zero in all amino-acid conditions.

Post-swap refinement also generally improved nucleotide reconstruction, but its effect was more dependent on the affinity representation and branch-length setting.
For random-tree datasets, mean paired differences ranged from $-0.6$ to 10.9 percentage points.
For Yule-tree datasets, they ranged from 0.1 to 8.7 percentage points.
JC69-local showed consistently positive improvements across all ten tree-generation and branch-length conditions, with mean differences ranging from 3.7 to 9.5 percentage points for random trees and from 4.4 to 7.5 percentage points for Yule trees.

Log-local also showed positive mean differences in all nucleotide conditions, with bootstrap confidence intervals above zero throughout.
The effects of refinement were less consistent for NormBit and Log-global at intermediate and longer branch lengths.
For Log-global in the random-tree datasets at a target mean branch length of 0.625, mean recovery was slightly lower after refinement, although the confidence interval included zero.
Several other NormBit and Log-global conditions also had confidence intervals that included zero.

\begin{figure*}[t]
    \centering

    \includegraphics[width=0.485\textwidth]
    {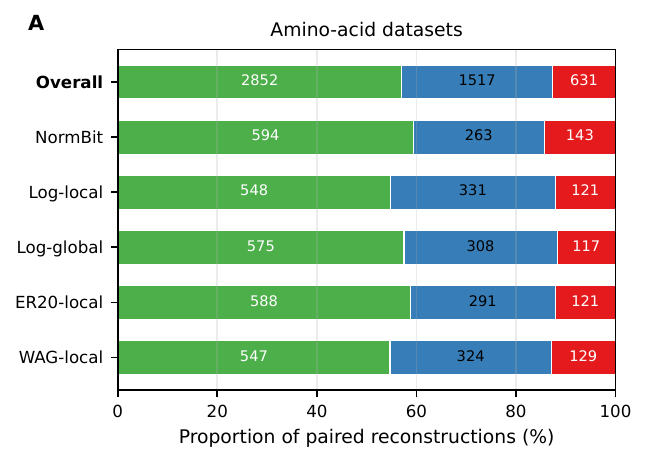}
    \hfill
    \includegraphics[width=0.485\textwidth]
    {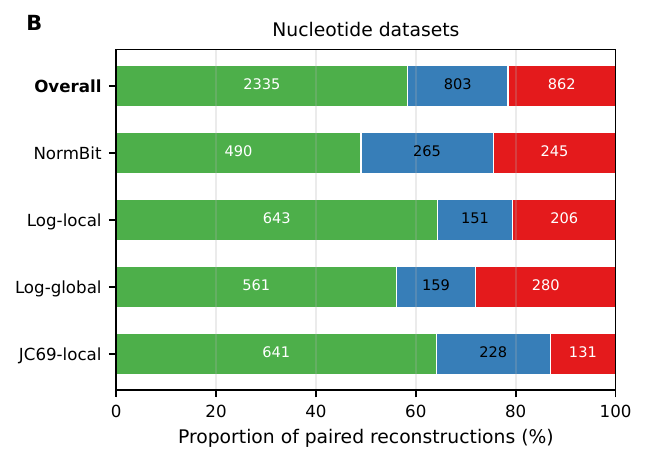}

    \par\vspace{-0.3em}

    \includegraphics[width=0.58\textwidth]
    {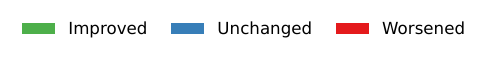}

    \caption{
    Reconstruction-level effects of post-swap refinement on internal
    split recovery.
    \textbf{(A)} Amino-acid datasets.
    \textbf{(B)} Nucleotide datasets.
    For each simulated dataset and affinity representation, the outcome was classified as improved, unchanged, or worsened according to whether split recovery with post-swap refinement was higher than, equal to, or lower than recovery without refinement.
    Bars indicate the proportions of paired reconstructions in each category, and numbers within the bars indicate counts.
    Each affinity representation comprised 1,000 paired reconstructions; the overall summaries comprised 5,000 amino-acid and 4,000 nucleotide reconstructions.
    All paired reconstructions were included; cases in which no post-swap exchange was accepted contributed to the unchanged category.
    }
    \label{fig:postswap_outcomes}
\end{figure*}

To examine whether the predominantly positive mean effects reflected consistent improvement across individual datasets, each paired reconstruction was classified as improved, unchanged, or worsened according to the change in split recovery after post-swap refinement (Fig.~\ref{fig:postswap_outcomes}).
Across the 5,000 paired amino-acid reconstructions, recovery improved in 2,852 cases (57.0\%), remained unchanged in 1,517 cases (30.3\%), and worsened in 631 cases (12.6\%).
Across the 4,000 paired nucleotide reconstructions, recovery improved in 2,335 cases (58.4\%), remained unchanged in 803 cases (20.1\%), and worsened in 862 cases (21.6\%).
Improved outcomes were more frequent than worsened outcomes for every affinity representation.
The proportion of worsened outcomes ranged from 11.7\% to 14.3\% among the amino-acid representations and from 13.1\% to 28.0\% among the nucleotide representations.

Overall, post-swap refinement more frequently improved than worsened split recovery, but its effect was not uniform across individual datasets or affinity representations.
By construction, each accepted post-swap exchange reduced the Ncut value of the candidate bipartition being refined.
The occurrence of worsened outcomes therefore shows that local improvement under the graph objective did not guarantee higher split recovery for the completed tree.
Worsened outcomes were less frequent in amino-acid reconstructions and in nucleotide reconstructions using JC69-local, but more frequent in nucleotide reconstructions using NormBit or BLAST-derived affinities.

\FloatBarrier

\subsection{Relative performance against neighbor joining depended on the distance representation}
\begin{figure*}[t]
    \centering

    \includegraphics[width=0.485\textwidth]
    {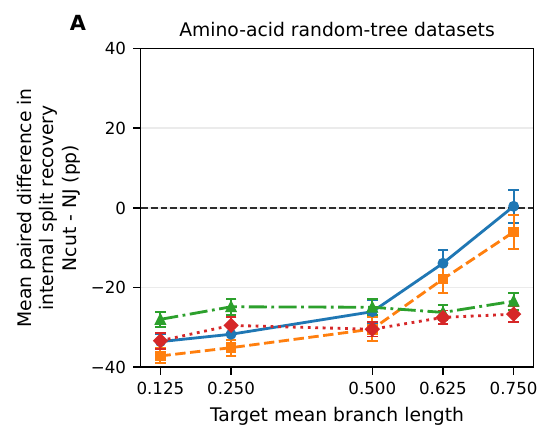}
    \hfill
    \includegraphics[width=0.485\textwidth]
    {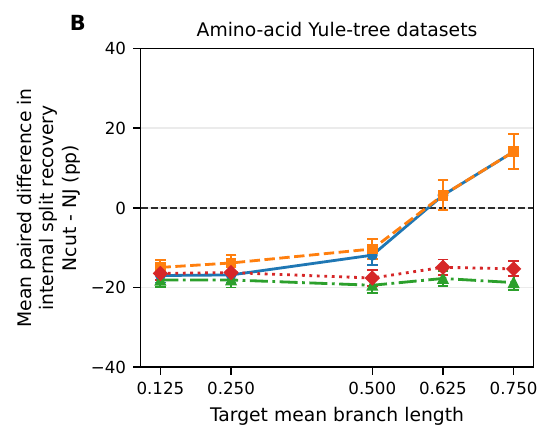}

    \par\vspace{0.3em}

    \includegraphics[width=0.485\textwidth]
    {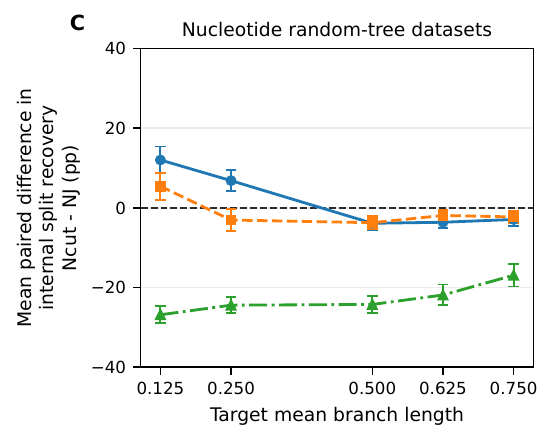}
    \hfill
    \includegraphics[width=0.485\textwidth]
    {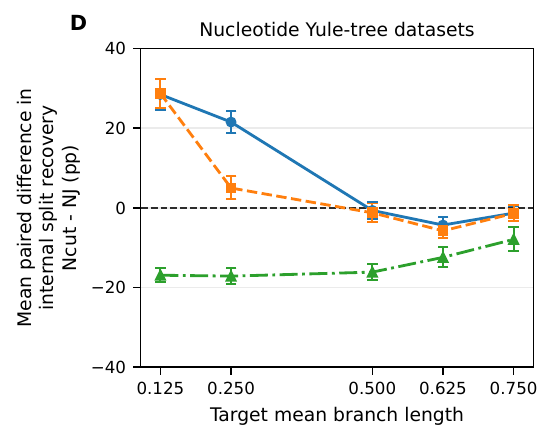}

    \par\vspace{-0.3em}

    \includegraphics[width=0.70\textwidth]
    {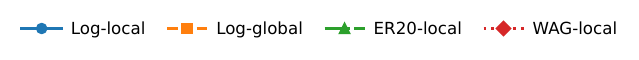}

    \par\vspace{-0.2em}

    \includegraphics[width=0.56\textwidth]
    {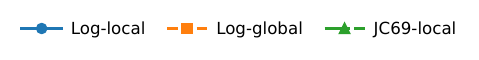}

    \caption{
    Differences in tree-reconstruction accuracy between recursive Ncut
    and neighbor joining.
    \textbf{(A)} Amino-acid random-tree datasets.
    \textbf{(B)} Amino-acid Yule-tree datasets.
    \textbf{(C)} Nucleotide random-tree datasets.
    \textbf{(D)} Nucleotide Yule-tree datasets.
    Recursive Ncut results were obtained without post-swap refinement.
    Differences were calculated as split recovery obtained by recursive Ncut minus that obtained by NJ for the same simulated dataset and underlying pairwise distance representation.
    Points indicate mean paired differences across 100 paired reconstructions per condition, and error bars indicate nonparametric bootstrap 95\% confidence intervals.
    Positive values indicate higher recovery by recursive Ncut, whereas negative values indicate higher recovery by NJ.
    }
    \label{fig:ncut_nj_condition}
\end{figure*}

\begin{figure*}[t]
    \centering

    \includegraphics[width=0.92\textwidth]
    {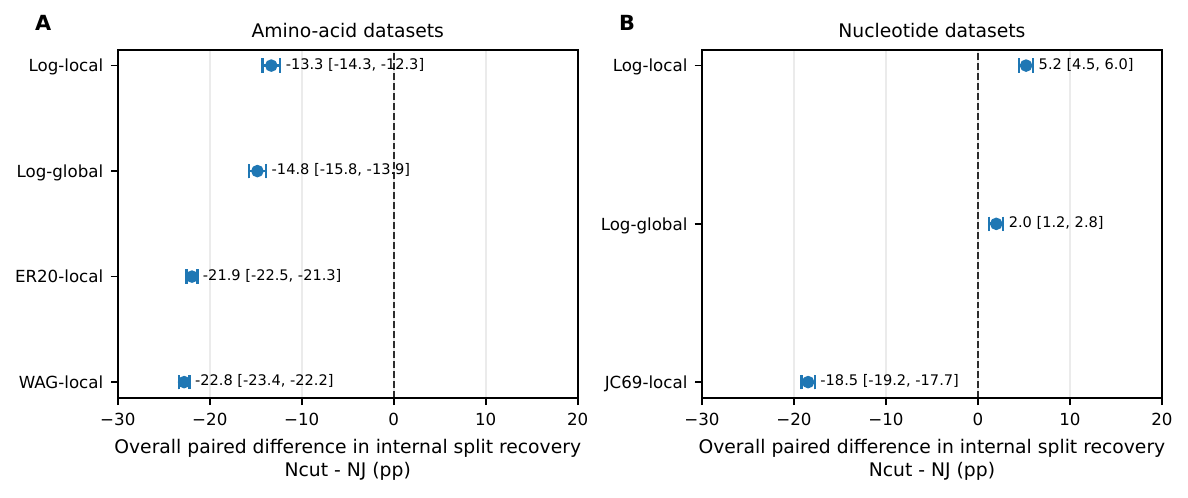}

    \caption{
    Overall paired differences in split recovery between recursive Ncut and neighbor joining.
    Differences were calculated as recursive Ncut minus NJ and were summarized across the two tree-generation settings and five target mean branch lengths.
    Bootstrap resampling was performed separately within each tree-generation-setting-by-branch-length stratum, and stratum means were averaged with equal weight.
    Each point summarizes 1,000 paired reconstructions for the corresponding distance comparison.
    Points indicate mean paired differences, and error bars indicate nonparametric bootstrap 95\% confidence intervals.
    Numerical annotations report the mean and 95\% confidence interval.
    Positive values favor recursive Ncut, whereas negative values favor NJ.
    }
    \label{fig:ncut_nj_overall}
\end{figure*}

We compared NJ, which reconstructs trees directly from pairwise distance matrices, with recursive Ncut, which uses affinity matrices derived from the corresponding pairwise distances (Figs.~\ref{fig:ncut_nj_condition} and~\ref{fig:ncut_nj_overall}).

For amino-acid datasets, WAG-local showed the highest overall mean recovery within the recursive Ncut framework.
However, NJ applied directly to the WAG distance matrices yielded higher mean recovery than the corresponding affinity-based recursive Ncut reconstruction in all tree-generation and branch-length conditions.
The overall mean paired differences, defined as recursive Ncut minus NJ, were $-21.9$ percentage points for ER20-local (95\% CI, $-22.5$ to $-21.3$) and $-22.8$ percentage points for WAG-local (95\% CI, $-23.4$ to $-22.2$).

Comparisons based on the BLAST-derived logarithmic distance also favored NJ overall.
The overall mean paired differences were $-13.3$ percentage points for Log-local (95\% CI, $-14.3$ to $-12.3$) and $-14.8$ percentage points for Log-global (95\% CI, $-15.8$ to $-13.9$).
This relationship was nevertheless condition dependent.
Recursive Ncut yielded higher mean recovery for both logarithmic-distance representations in the Yule-tree datasets at target mean branch lengths of 0.625 and 0.750.
Log-local also yielded a small positive mean difference in the random-tree datasets at a target mean branch length of 0.750.

For nucleotide datasets, JC69-local achieved the highest recovery among the affinity representations evaluated within recursive Ncut.
Nevertheless, NJ applied directly to the same JC69 distance matrices yielded higher mean recovery than the corresponding affinity-based recursive Ncut reconstruction in all tree-generation and branch-length conditions.
The overall mean paired difference was $-18.5$ percentage points (95\% CI, $-19.2$ to $-17.7$).

In contrast, comparisons based on the BLAST-derived logarithmic distance favored recursive Ncut overall.
The overall mean paired differences were 5.2 percentage points for Log-local (95\% CI, 4.5--6.0) and 2.0 percentage points for Log-global (95\% CI, 1.2--2.8).
These advantages were concentrated at shorter target mean branch lengths.
At target mean branch lengths from 0.500 to 0.750, all condition-level mean differences were negative and therefore favored NJ, although their magnitudes were generally small.

The affinity representations that performed best within the recursive Ncut framework were not necessarily those for which recursive Ncut performed best relative to NJ.
NJ showed a substantial and consistent advantage over affinity-based recursive Ncut when ER20, WAG, or JC69 distances were used.
For the BLAST-derived logarithmic distance, however, the relative performance of the two reconstruction approaches depended more strongly on sequence type, tree-generation setting, and evolutionary divergence.

\FloatBarrier

\section{Discussion}
Accurate phylogenetic reconstruction depends on how information about evolutionary relationships is extracted from sequence data and incorporated into a tree-building procedure.
Here, we examined how alternative representations of pairwise sequence relationships affect tree reconstruction within a recursive graph-partitioning framework.
Graph construction affected recursive Ncut-based phylogenetic reconstruction, with the clearest differences observed in nucleotide datasets.
JC69-local retained comparatively high split recovery as evolutionary divergence increased, whereas the BLAST-derived representations declined more markedly; differences among amino-acid representations were smaller and more condition dependent.
Comparison with NJ further showed that pairwise distances yielding strong performance after affinity transformation within recursive Ncut could nevertheless produce more accurate trees when used directly by NJ.

Affinity construction should therefore be regarded as an integral component of recursive Ncut reconstruction rather than as neutral preprocessing.
Because Ncut evaluates partitions relative to the supplied edge weights, its performance depends on how those weights represent phylogenetic proximity~\cite{Shi2000,Maier2013}.
Neither local scaling nor model correspondence alone explained the observed patterns: JC69-local may have benefited from partial model agreement, whereas WAG-local was not consistently superior despite WAG-based simulation, and neither distance calculation incorporated the simulated gamma-rate heterogeneity.

Comparison with NJ showed that reconstruction accuracy depends not only on the pairwise distance representation, but also on how that information is used to construct the tree.
Although WAG-local and JC69-local performed well within recursive Ncut, NJ achieved higher split recovery when the corresponding distances were used directly.
This indicates that evolutionary distances that perform well after affinity transformation within recursive Ncut may still be used more effectively by a direct distance-based method such as NJ.
However, because the present design cannot separate the effects of affinity transformation from those of recursive partitioning, the source of this difference remains unclear.
Conversely, recursive Ncut outperformed NJ for BLAST-derived logarithmic distances under some conditions, suggesting that the relative performance of the two reconstruction approaches depends on how the underlying sequence relationships are represented.

Post-swap refinement directly illustrated the mismatch between optimization of an internal graph objective and recovery of the reference topology.
Although each accepted swap reduced Ncut, worsened split recovery occurred for every affinity representation; moreover, modifying one bipartition changes the subsets passed to subsequent recursive steps.
Thus, local Ncut improvement does not guarantee improved tree accuracy, consistent with NMcutDA~\cite{Onodera2023} and with reports that heuristic graph-splitting optimization can occasionally yield higher branch recovery than exact-solver NMcut reconstruction~\cite{FernandezOtero2026}.

Several limitations should be considered.
The study was restricted to simulated alignments of 30 taxa and 500 sites under a limited set of evolutionary conditions, with fixed kernel and optimization parameters.
The NJ comparison evaluated complete reconstruction pathways and therefore could not separate affinity conversion from recursive partitioning.
Empirical evaluation will also be necessary to determine whether affinity-based reconstruction can usefully integrate sequence, structural, functional, or genomic information under more complex evolutionary conditions.
Finally, although Simulated Bifurcation is amenable to parallel computation~\cite{Goto2021}, repeated optimization across cardinalities and recursive nodes means that solver-level scalability does not establish scalability of the full reconstruction procedure.

Overall, graph construction influenced recursive Ncut-based phylogenetic reconstruction, but improvements in affinity representation did not ensure superiority over direct distance-based reconstruction, highlighting the need for joint evaluation of pairwise representation, affinity transformation, optimization, and recursive tree construction on larger simulated and empirical datasets.
More broadly, determining how sequence-derived evolutionary information should be represented may help clarify when graph-based reconstruction can provide useful alternatives or complements to established phylogenetic methods.

\section*{Code and data availability}
The source code and released data supporting this study are available at \url{https://github.com/Keio-University-Bio2Q/graph-construction-in-qubo-based-recursive-phylogenetic-tree-reconstruction}. The repository contains the amino-acid and nucleotide analysis notebooks, simulated reference trees, simulated sequence datasets, normalized public-release manifests, neighbor-joining input distance matrices, and the reference evaluation tables used for the reported aggregate analyses. Simulation seeds are included in the released manifests. Methodological details and software versions are described in the Materials and Methods section.

\section*{Acknowledgements}
Human Biology-Microbiome-Quantum Research Center (Bio2Q) is supported by World Premier International Research Center Initiative (WPI), MEXT, Japan. This work was also supported by the Center of Innovation for Sustainable Quantum AI (JST Grant Number JPMJPF2221). The authors thank Ashish Joshi for helpful comments on the manuscript and Scott Behie for valuable discussions and comments on the manuscript.

\bibliographystyle{unsrt}
\bibliography{ref}

\end{document}